\documentclass[aps,prl,multicol,groupedaddress]{revtex4}
\usepackage{graphicx}\usepackage{amsmath}\usepackage{amssymb}
\begin{document}
\title{\bf  Mixing and decoherence to nearest separable states in quantum measurements}
\author{Avijit Lahiri}
\email{a_l@vsnl.com}
\affiliation{ Dept of Physics, Vidyasagar Evening College, Kolkata 700 006, INDIA}


\begin{abstract}
We illustrate through numerical results a number of features of environment-induced decoherence under a broad class of apparatus-environment interactions in quantum measurements wherein the reduced system-apparatus density matrix evolves towards the nearest separable state and, in addition, there occurs a mixing in relevant groups of apparatus microstates (see below). The resulting final state is unique and correctly embodies the measurement statistics even in the absence of environment-induced superselection because of energy differences between these groups of states. The partial transpose remains non-positive throughout the process.
\end{abstract}
\pacs{03.65.Ta, 03.67.-a}
\maketitle

\noindent The measurement problem has traditionally posed questions in quantum theory relating to interpretation at a basic level. More recently, however, investigations on the process of environment-induced decoherence seem to have provided some of the answers, at least as plausible explanations. The present paper aims at illustrating in simple and general terms a number of features of this process whereby entangled states of a certain type (to be termed pure-mixed entanglement, see below) involving the measured system and the apparatus go through a Brownian-like evolution in the composite state space, tending towards the nearest separable state (in the sense explained in~\cite{vedral-plenio}) with an attendant erosion of the entanglement and, additionally, there occurs a mixing in the apparatus states leading to a homogeneous probability distribution over relevant sets of microstates of the latter. The mixing causes a partial erosion in classical correlations as well, but at the end the remaining classical correlation between the measured system and the apparatus correctly describes the possible results of measurement and their respective probabilities. Moreover, for the class of entangled states considered, non-positivity of the partial transpose seems to be retained throughout the decoherence process, thereby providing a convenient indicator for monitoring the entire process.
\vskip .5cm
\noindent
More precisely, we consider three systems that interact with one another and that can be interpreted, in an appropriate context, as a measured system (S), a measuring apparatus (A), and an environment (E). The state spaces of the three systems have, in general, different dimensions - a fact of crucial importance in the context of the quantum measurement problem. For the composite system SA, we consider a situation where pure states of S are entangled with mixed states of A, which we refer to as pure-mixed entanglement, and which arises naturally from pure-pure entanglement by environment-induced dephasing among sets of apparatus states (see below) during the process of decoherence. The former has been considered in~\cite{vedral} in relation to the information gain in a quantum measurement. 
\vskip .5cm
\noindent
Our numerical results refer to a 2D system S with orthonormal states $|s_1>,~|s_2>$ (eigenstates of the observable to be measured), an apparatus A with two bunches of orthonormal states (say, $|a_1>,\ldots ,|a_{N_1}>,~|b_1>,\ldots ,|b_{N_2}>$) forming subspaces of dimensions $N_1,~N_2$, and an environment E with a state space of dimension $N_e$. The bunches of apparatus states are almost-degenerate microstates corresponding to macroscopic pointer states that may be superpositions or mixtures of these microstates (see, for instance,~\cite{vedral,peres1}). Mixed apparatus states are actually more relevant in the context of measurement~\cite{wigner}, since the apparatus is distinct from the measured system in that it is a macroscopic system for which mixed states arise naturally due to environmental perturbations (see, e.g.,~\cite{peres1}).
\vskip .5cm
\noindent
The S-A state we start from is given by

\begin{eqnarray}
\rho=|c_1|^2|s_1><s_1|\otimes \rho^{(A)}_a + |c_2|^2|s_2><s_2|\otimes \rho^{(A)}_b~~~~~~~~~~~~~~~~~~~~~~~~~~ \nonumber\\~~~~~~~+c_1c_2^*|s_1><s_2|\otimes |\phi^{(A)}_a><\phi^{(A)}_b|+c_1^*c_2|s_2><s_1|\otimes |\phi^{(A)}_b><\phi^{(A)}_a|, \label{1}
\end{eqnarray}

\noindent where $c_1,~c_2$ are superposition coefficients of the system state ($c_1|s_1>+c_2|s_2>$) being measured,
\begin{subequations}
\begin{equation}
\rho^{(A)}_a\equiv \sum_{i=1}^{N_1}p_i|a_i><a_i|,\label{2a}
\end{equation}

\noindent and

\begin{equation}
\rho^{(A)}_b\equiv \sum_{i=1}^{N_2}q_i|b_i><b_i|, \label{2b}
\end{equation}
\end{subequations}

\noindent are the mixed apparatus states entangled with $|s_1>$, and $|s_2>$ respectively, and 
\begin{subequations}
\begin{equation}
|\phi^{(A)}_a>\equiv \sum_{i=1}^{N_1}p_i|a_i>, \label{3a}
\end{equation}
\begin{equation}
|\phi^{(A)}_b>\equiv \sum	_{i=1}^{N_2}q_i|b_i>. \label{3b}
\end{equation}
\end{subequations}

\noindent are involved in the off-diagonal terms representing the entangled state. 

\vskip .5cm
\noindent
In these expressions, $\{p_i\},~ \{q_i\}$ are two sets of weights specifying the degree of mixing in the two groups of apparatus states. A pure apparatus state corresponds to one of the relevant set of weights being unity, with all the other weights of the set being zero. The commonly discussed special case of pure-pure entanglement corresponds to both the apparatus states being pure (and so, $\rho^{(A)}_a=|\phi^{(A)}_a><\phi^{(A)}_a|, ~\rho^{(A)}_b=|\phi^{(A)}_b><\phi^{(A)}_b|$), though in reality such a state is transformed into a state of the form (\ref{1}) by environmental dephasing referred to above. Various different pure-mixed entangled states with a given degree of mixing can be constructed by invoking an auxiliary system purifying the apparatus state~\cite{vedral} and then tracing out this auxiliary system, but we use (\ref{1}), which is symmetric in the two sets of apparatus states, without loss of generality. Note that while $|\phi^{(A)}_a>,~|\phi^{(A)}_b>$ have norms, in general, different from unity, (\ref{1}) happens to be satisfy all the requirements of a density matrix. 
\vskip .5cm
\noindent
The pure-mixed entangled state (\ref{1}) has a structure similar to that of a pure-pure one of the form
\begin{eqnarray}
\rho^{(P-P)}=|c_1|^2|s_1><s_1|\otimes |\phi^{(A)}_1><\phi^{(A)}_1|+|c_2|^2|s_2><s_2|\otimes |\phi^{(A)}_2><\phi^{(A)}_2|\nonumber\\+
c_1c_2^*|s_1><s_2|\otimes |\phi^{(A)}_1><\phi^{(A)}_2|+
c_1^*c_2|s_2><s_1|\otimes |\phi^{(A)}_2><\phi^{(A)}_1|. \label{4}
\end{eqnarray}

\noindent
The end result  arising from the environment-induced decoherence operating on (\ref{4}) and embodying correctly the measurement statistics (relating to the measurement variable under consideration, represented by $s_1|s_1><s_1|+s_2|s_2><s_2|$, where $s_1,~s_2$ are the relevant eigenvalues) is

\begin{equation}
\sigma^{(P-P)}=|c_1|^2|s_1><s_1|\otimes |\phi^{(A)}_1><\phi^{(A)}_1|+|c_2|^2|s_2><s_2|\otimes |\phi^{(A)}_2><\phi^{(A)}_2|, \label {5}
\end{equation}

\noindent
obtained by deleting the off-diagonal terms, since one can assume that the latter are averaged out due to the decoherence. In a similar vein, one expects that the disentangled state resulting from (\ref{1}) would be

\begin{equation}
\sigma^{(P-M)}=|c_1|^2|s_1><s_1|\otimes \rho^{(A)}_a + |c_2|^2|s_2><s_2|\otimes \rho^{(A)}_b.\label{6}
\end{equation}

\noindent
While the state (\ref{5}) has the interesting property that it is the closest separable state to (\ref{4}), in the sense explained in~\cite{vedral-plenio}, we indicate numerical evidence below to show that (\ref{6}) is similarly the nearest separable state to (\ref{1}). Thus, a neat description of the process of environment-induced decoherence would be that it is a quantum operation transforming the initial S-A entangled state to the nearest separable state. We see in our numerical results below that the decoherence process under quite general A-E interactions indeed shows a tendency towards such a state, but additionally involves another transformation in the apparatus states, namely one towards maximum mixing in the two groups of states $|a_1>,\ldots ,|a_{N_1}>$, and $|b_1>,\ldots ,|b_{N_2}>$ leading us to consider the separable state 

\noindent
\begin{equation}
\sigma^{(P-M)}_0=|c_1|^2|s_1><s_1|\otimes \sigma^{(A)}_a + |c_2|^2|s_2><s_2|\otimes \sigma^{(A)}_b,\label{7}
\end{equation}

\noindent
where $\sigma^{(A)}_a$ and $\sigma^{(A)}_b$ correspond to microcanonical ensembles in the two groups of apparatus states:
\begin{subequations}
\begin{equation}
\sigma^{(A)}_a=\frac{1}{N_1}\sum_{i=1}^{N_1}|a_i><a_i|,\label{8a}
\end{equation}
\begin{equation}
\sigma^{(A)}_b=\frac{1}{N_2}\sum_{i=1}^{N_2}|b_i><b_i|.\label{8b}
\end{equation}
\end{subequations}

\noindent
The total Hamiltonian we consider is given by (for the sake of generality, the dimensions of the state spaces of the three systems are denoted by $N_s, N_a$, and and $N_e$ respectively; as indicated above, we use $N_s=2$, $N_a=N_1+N_2$, with appropriate choices for $N_1,~N_2$, as also for $N_e$, see below):

\begin{equation}
H= I_S\otimes H_A\otimes I_E+I_S\otimes I_A\otimes H_
E+\lambda I_S\otimes V_{A-E}   \label{9}
\end{equation}

\noindent
Here $I_S, I_A, I_E$ denote identity operators for S, A, and E respectively, $ H_A,~ H_E$ represent the Hamiltonians for A and E considered in isolation, diagonal in the respective sets of basis states chosen, and $V_{A-E}$ stands for the A-E interaction responsible for the decoherence, with strength $\lambda$.

\vskip .5cm
\noindent
Before presenting our results, and at the cost of some repetition, we summarise the basic assumptions underlying the present paper:

\vskip .5cm
\noindent
(i) the S-A interaction is not included in (\ref{9}) where it is assumed that its effect is described by the S-A entangled state (\ref{1}) on which (\ref{9}) acts; this is the so-called pre-measurement unitary evolution;
\vskip .5cm
\noindent
(ii) no S-E interaction term is included in (\ref{9}); while the decohering effect of E is expected to act on S as on A, the typical decoherence time is expected to be large since S is supposed to be a small quantum system; in any case we assume that this interaction is not of relevance in our present context;
\vskip .5cm
\noindent
(iii) in other words there are several time scales involved in a typical quantum measurement; the S-E interaction takes the longest to have its effect felt; the S-A measurement intearction seemingly goes on along with the dephasing effect of E on A, the two together generating the S-A entangled state with pure states of S entangled with mixed states of A; however, (1) includes the possibility of pure-pure entanglement as a special case and hence the assumption of an initial pure-mixed entangled state is not a restrictive one;
\vskip .5cm
\noindent
(iv) the A-E interaction disentangles the S-states from the A-states, leading to a state in which only the classical correlations between the two remain; for such a state one can talk in terms of 'pre-existing' properties in S; on a longer time scale, the environment can be expected to destroy the classical correlations as well, dissipating all the information inherent in the initial S-A state;
\vskip .5cm
\noindent
(v) the free evolution os S is not included separately in (9) since the quantum nature of S is expected to imply only a small energy difference between $|s_1>$ and $|s_2>$ as compared with the relevant energy differences for the apparatus and the environment states; in other words, in the context of the process of decoherence, the system states are assumed to be energetically degenerate;
\vskip .5cm
\noindent
(vi) finally, the A-E interaction is assumed to be such as not to cause significant mixing between the two groups of states $|a_1>,\ldots ,|a_{N_1}>$ and $|b_1>,\ldots ,|b_{N_2}>$; in other words, the decoherence process preserves the pointer variables pertaining to the two groups of apparatus states, in terms of which values are assigned to the observable of S being measured; while this can be assumed to be a consequence of a symmetry inherent in the A-E interaction, responsible for what has been termed environmental selection or 'einselection'~\cite{zurek,schlosshauer}, it can equally well be explained as a consequence of the energy difference between the two groups of apparatus states, the latter being large enough to effectively suppress mixing between the two; in simple terms, the macroscopic pointer states are stable against environmental perturbations, at least on the time scale on which the decoherence acts, even without any special symmetry in the A-E interaction being responsible for it.
\vskip .5cm
\noindent
This brings us to the choice for the A-E interaction operator $V_{A-E}$. Since the apparatus and the environment are both macroscopic systems with many degrees of freedom, the most realistic choice should be an operator represented by a random Hermitian matrix in an arbitrarily chosen basis; indeed, any other form would imply some special assumption or other relating to the interaction and would be contrary to the macroscopic nature of the two systems in interaction. The latter are effectively classical ones~\cite{peres1} with densely bunched degenerate states, whose interactions are, generically speaking, chaotic in the classical description. The quantum features of such interactions are known to be similar to those of ensembles of random matrices. A large body of recent work has looked into the entangling power of chaotic interactions (see, e.g.,~\cite{lakshminarayan,scott,sanders}), and a number of these also bring out random features in the density matrix fluctuations in subsystems interacting with one another through such random matrices~\cite{lahiri,nag-ghosh}.

\vskip .5cm
\noindent
The above A-E interaction efficaciously entangles the apparaus states with the environment states and at the same time disentangles the system states from the apparatus states(see, e.g.,~\cite{wooters}, for an introduction to entanglement sharing in tripartite systems). The reduced S-A density matrix elements fluctuate during the process, whereby $\rho_{S-A}$ undergoes a Brownian-like motion in the space of entangled states, tending to the nearest separable state (\ref{6}) while at the same time deviating from the latter due to the mixing among the two groups of apparatus states alluded to above, finally reaching the stae (\ref{7}).

\vskip .5cm
\noindent
The whole trajectory of the S-A state can be conveniently monitored through the partial transpose (with respect to, say, S) of $\rho_{S-A}$, defined as

\begin{equation}
<s_i\alpha|\rho^{T_S}_{S-A}|s_j\beta>=<s_j\alpha|\rho_{S-A}|s_i\beta>~~(i,j=1,2), \label{10}
\end{equation}
\begin{figure*}[ht]
	\centering
		\includegraphics[width=0.9\columnwidth]{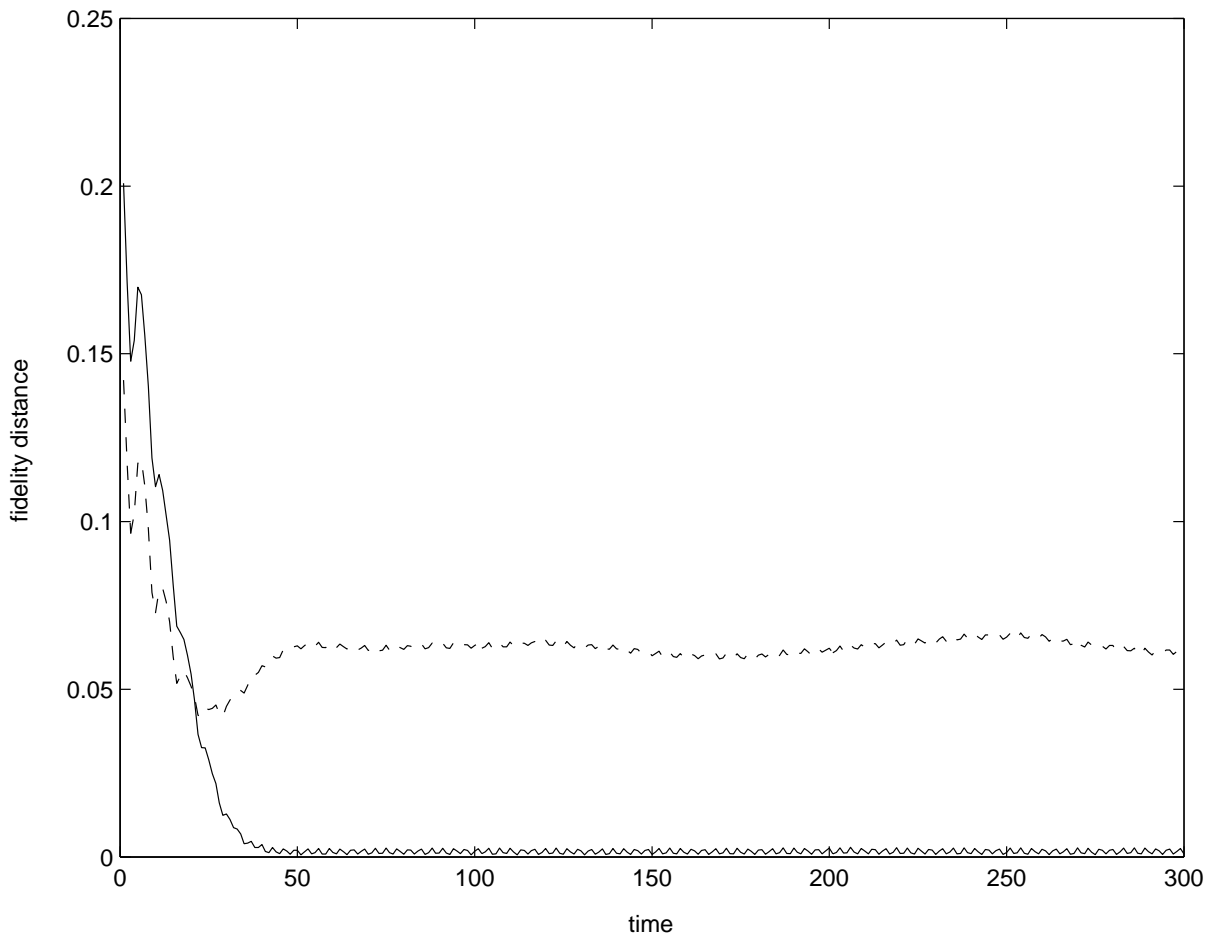}
	\caption{\label{cap:fig1}variation with time (arbitrary scale) of fidelity distance (see text) of the evolved reduced S-A state from the states (6) (dashed line) and (7) (solid line); $N_1=3,~N_2=4,~ N_e=22,~ \lambda=0.04$, $c_1=c_2=\frac{1}{\sqrt{2}}$; the weights $\{p_i\},~\{q_i\}$ in the initial state (\ref{1}) are chosen randomly; the environment state is taken as a uniform mixture of basic states, the latter covering a small energy band in the middle of the energy gap between the two groups of apparatus states; this is an artefact for preventing mixing between the two groups; more generally,  small values of $\lambda$ prevent the mixing within the relevant time scale.}
\end{figure*}

\noindent
where $|\alpha>,~|\beta>$ are any two of the basic apparatus states mentioned above. It is known that the partial transpose is a positive operator for separable states while, for an entangled state, it may or may not be positive. For the state (\ref{1}), however, it happens to be a non-positive operator, with one negative eigenvalve. For instance, with $N_1=1,~ N_2=2$, for which the basic apparatus states are, say, $|a>$ and $|b_1>,~|b_2>$, while $\rho^{(A)}_a$ and $\rho^{(A)}_b$ (refer to (\ref{2a}), (\ref{2b})) are given by $|a><a|$ and ($q_1|b_1><b_1|+q_2|b_2><b_2|$), the eigenvalues of the partial transpose of (\ref{1}) are $\pm |c_1c_2|\sqrt{q_1^2+q_2^2},~0,~|c_1|^2,~|c_2|^2q_1,~|c_2|^2q_2$ (in this case the system S-A is 2x3 dimensional, for which non-positivity of the partial transpose is known to be a necessary and sufficient condition for entanglement~\cite{horodecki}). Similar results  obtain for other values of $N_1,~N_2$ irrespective of the weights $\{p_i\},~\{q_i\}$. What is more, the partial transpose remains non-positive during the entire course of decoherence to (\ref{7}), as we see in our numerical results below.

\begin{figure*}[ht]
	\centering
		\includegraphics[width=0.9\columnwidth]{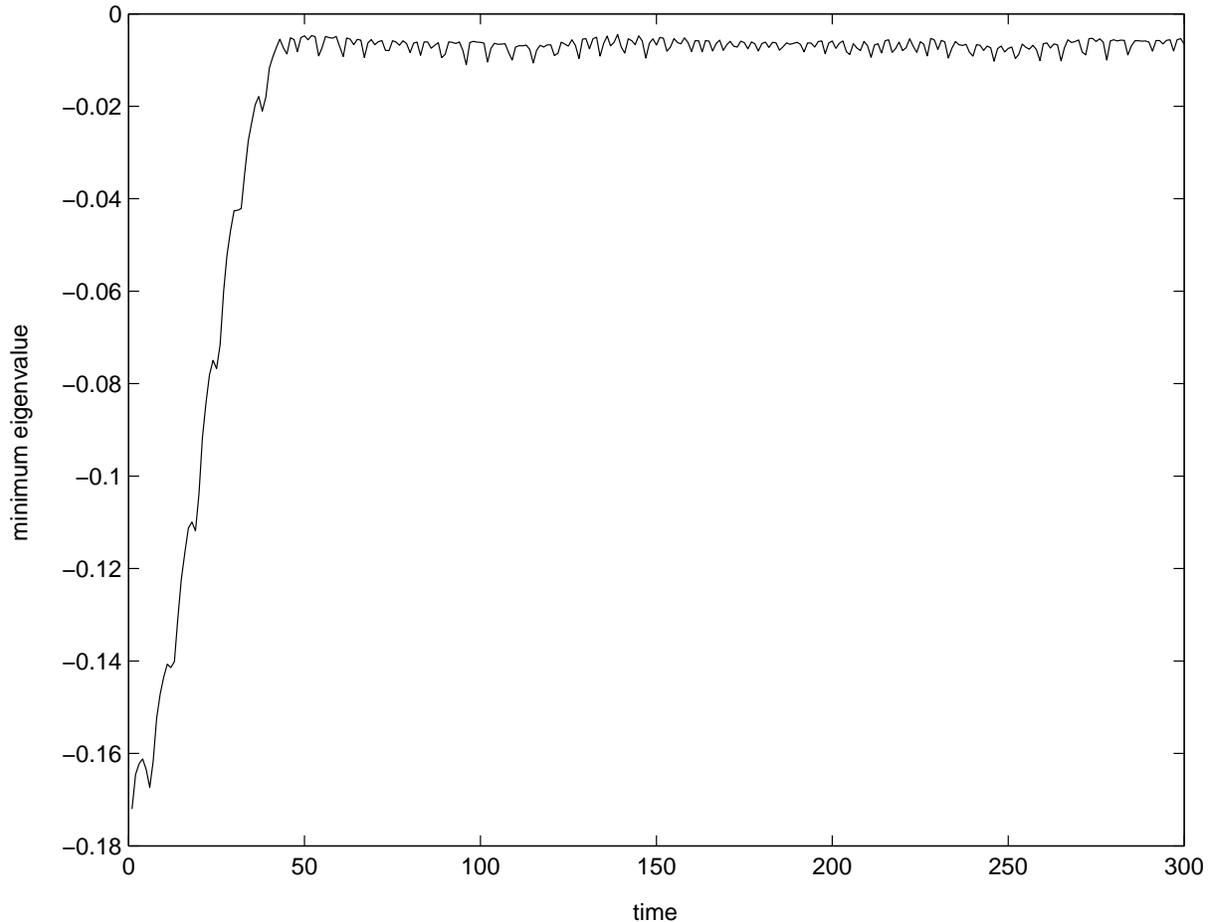}
	\caption{\label{cap:fig2} Variation with time of the lowest eigenvalue of the partial transpose (see text) of the evolved reduced S-A state; other particulars as in Fig.~\ref{cap:fig1}. }
	
\end{figure*}

\noindent Fig.~\ref{cap:fig1} (see caption) shows the course of the decoherence process by plotting the fidelity distance (defined as $(2-2 ~trace~(\rho^{1/2}\sigma\rho^{1/2})^{1/2})$ for any two density matrices $\rho,~\sigma$) of the reduced S-A state from the states (\ref{6}) (dashed line) and (\ref{7}) (solid line) against time (arbitrary units), where it is seen that while initially (\ref{7}) was at a greater distance as compared to (\ref{6}) from (\ref{1}) (since (\ref{6}) is actually the nearest separable state to (\ref{1}), see below) and while during the initial stage the evolved state does approach (\ref{6}), it eventually gets closer to and ends up at (\ref{7}), confirming the two principal trends at work referred to above. At the same time Fig.~\ref{cap:fig2} shows that the minimum eigenvalue of the partial transpose remains negative throughout the course of this evolution finally attaining the value zero.

\vskip .5cm
\noindent
We mention that a systematic numerical procedure generating convex combinations of arbitrarily chosen product states with (\ref{6}) indicates that (\ref{6}) is indeed the nearest separable state to (\ref{1}), as (\ref{5}) is to (\ref{4})(to be reported elsewhere).

\vskip .5cm
\noindent
These numerical results illustrate the principal features of the decoherence process outlined in the preceding paragraphs, namely, the three basic principles governing the evolution of the reduced S-A state during decoherence in a quantum measurement are (a) a tendency towards the nearest separable state, (b) a tendency towards homogeneous mixing between the groups of apparaus states corresponding to distinct values of the pointer variable, and (c) Brownian-like fluctuations in the reduced density matrix. More detailed results will be reported elsewhere. One significant fact to emerge is that the partial transpose remains non-positive throughout the process of decoherence.

\vskip .5cm
\noindent
The above features, moreover, arise under the most general randomly generated S-A interactions, even in the absence of special environmental variables and states involved in any specific type of entanglement with the environment, viz., one describing environmantal selection. The uniqueness of the final state resulting from the decoherence is a consequence of the energy differences between the groups of relevant apparatus states rather than of environmental selection.

\vskip .5cm
\noindent
It has been argued in~\cite{vedral} that the information gain in a quantum measurement is actually the same as the classical correlation beteen the system and the apparatus states, and by implication, the degree of quantum entanglement between the two is not relevant, the latter being precisely the information erased during the decoherence process. For a quantum system and an apparatus with a small and a large number of degrees of freedom respectively, the magnitude of quantum entanglement is actually small (see, for, instance~\cite{zyckowski}, results will also be presented in a separate communication) as compared with the distance from the initial entangled state to, say, faraway separable states. It is this smallness of the quantum information to be erased that is responsible for the process of 'seeking out', in a Brownian-like evolution, of the appropriate disentangled state correctly embodying the measurement statistics.

\vskip .5cm
\noindent
We have considered only projective measurements on pure system states in this paper. The results are seen to hold for measurements on mixed system states as well (to be reported). It is not, however, clear whether corresponding statements can be made for more general POVM operations as well. An application of the features observed in this paper in devising a general numerical procedure for evaluating the magnitude of bipartitie entanglement will be described elsewhere.
\vskip .5cm
\noindent {\bf ACKNOWLEDGEMENT:} Thanks are due to Gautam Ghosh, Saha Institute of Nuclear Physics, Kolkata, for helpful discussions.

\end {document}